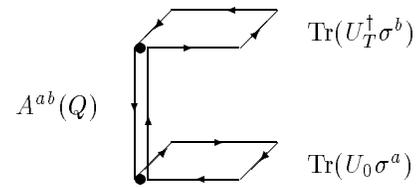

**FIGURE 1**

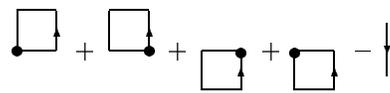

(a)

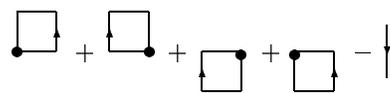

(b)

**FIGURE 2**

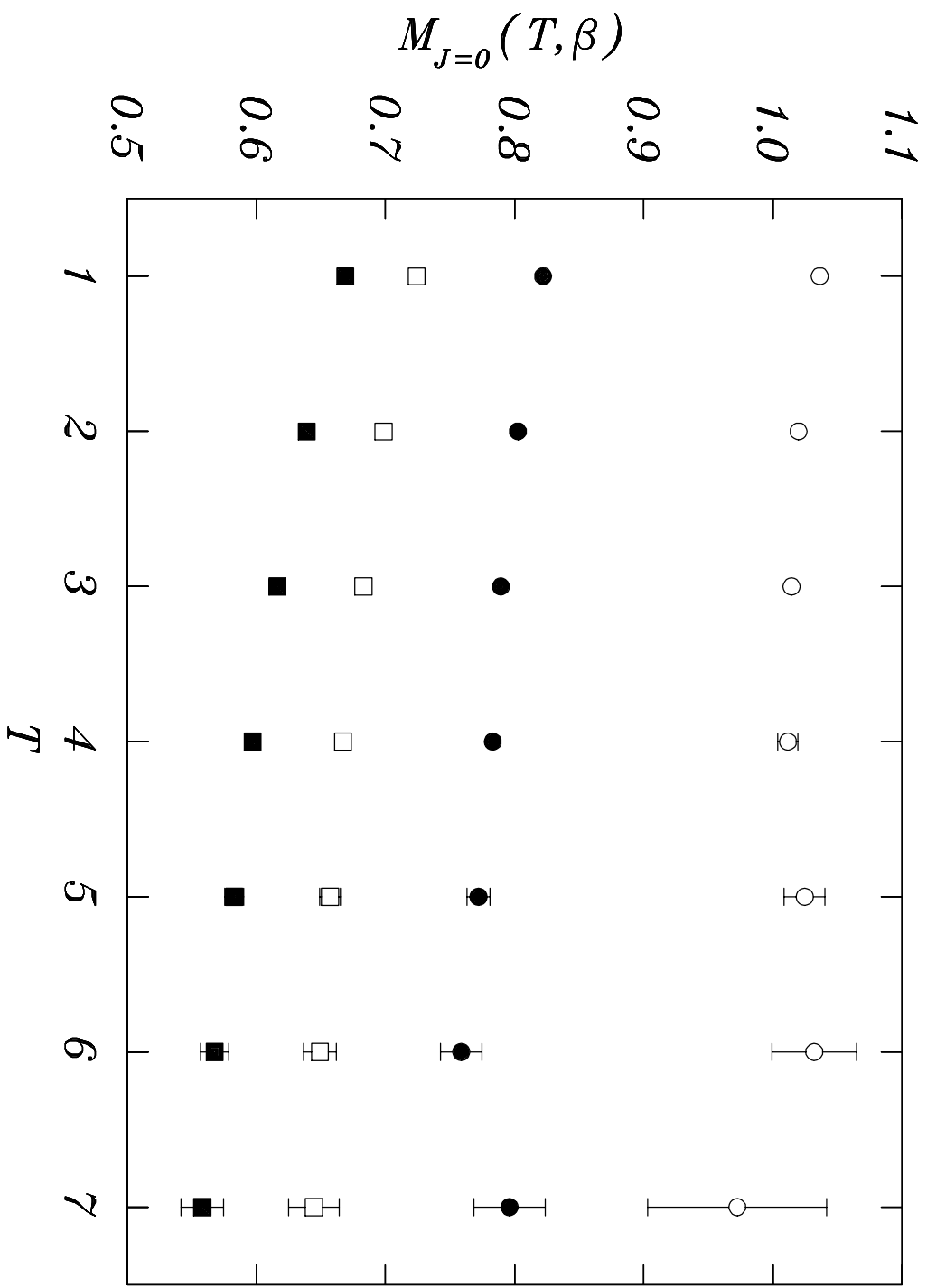

FIGURE 3

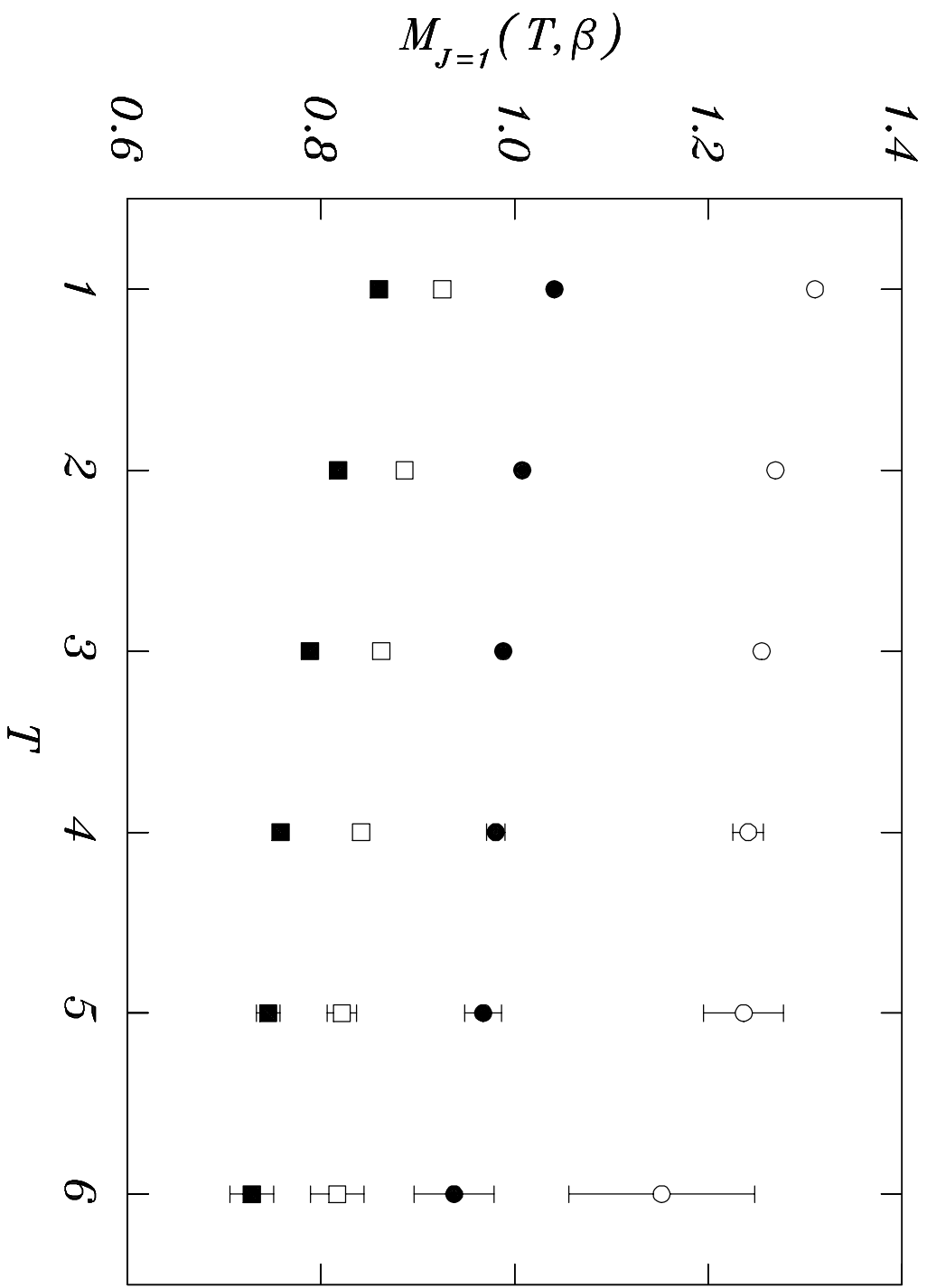

FIGURE 4

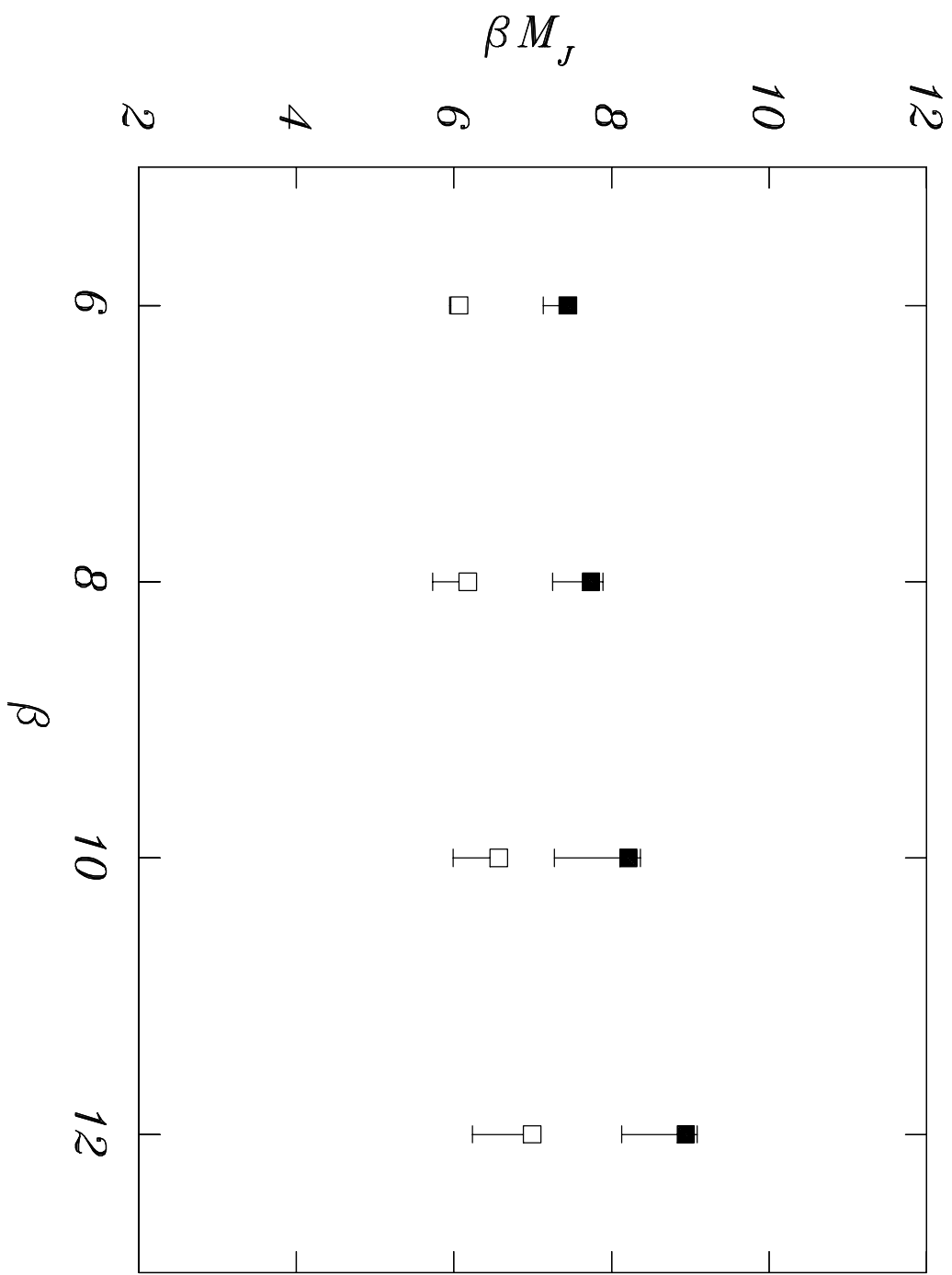

FIGURE 5

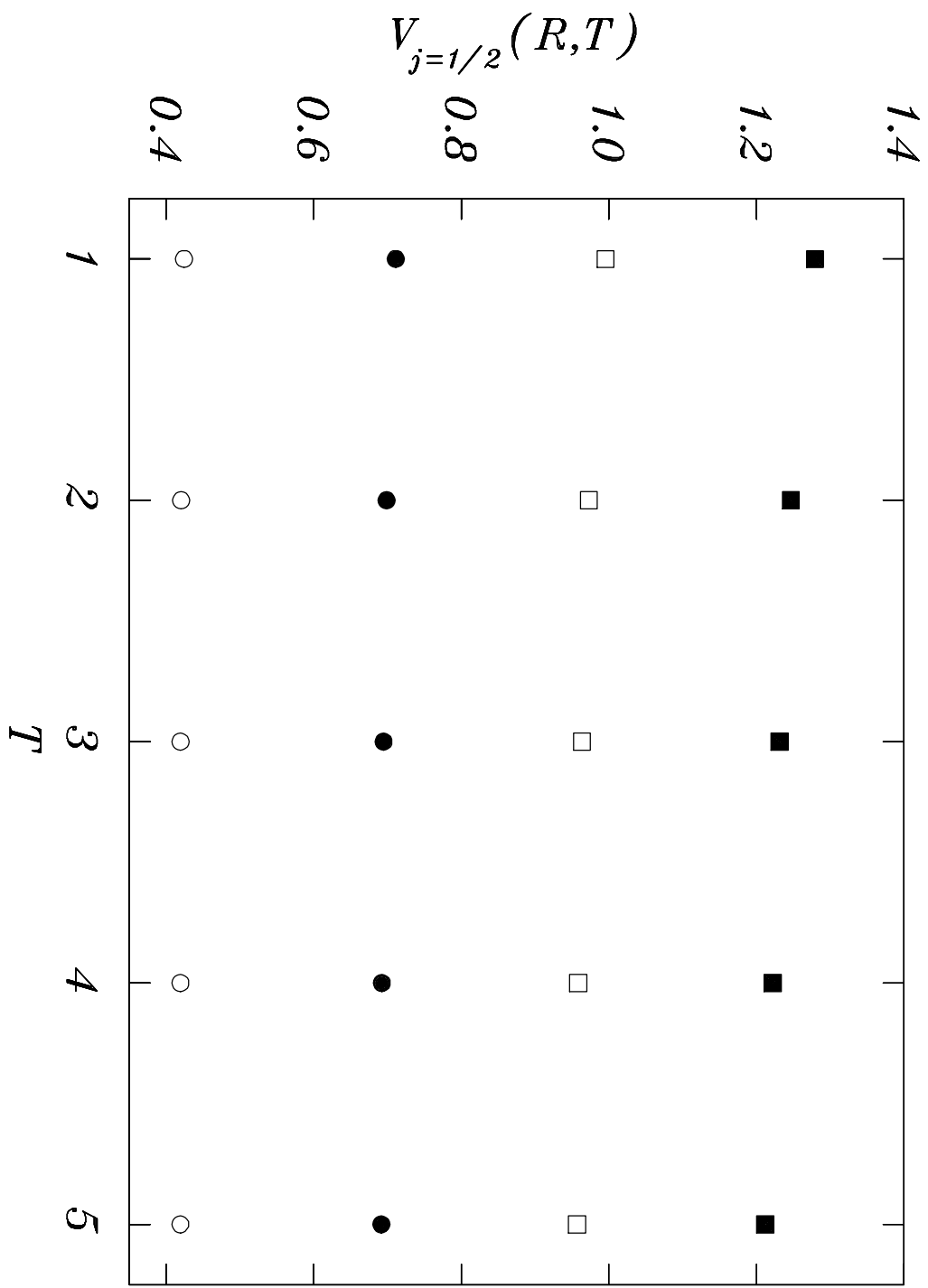

FIGURE 6

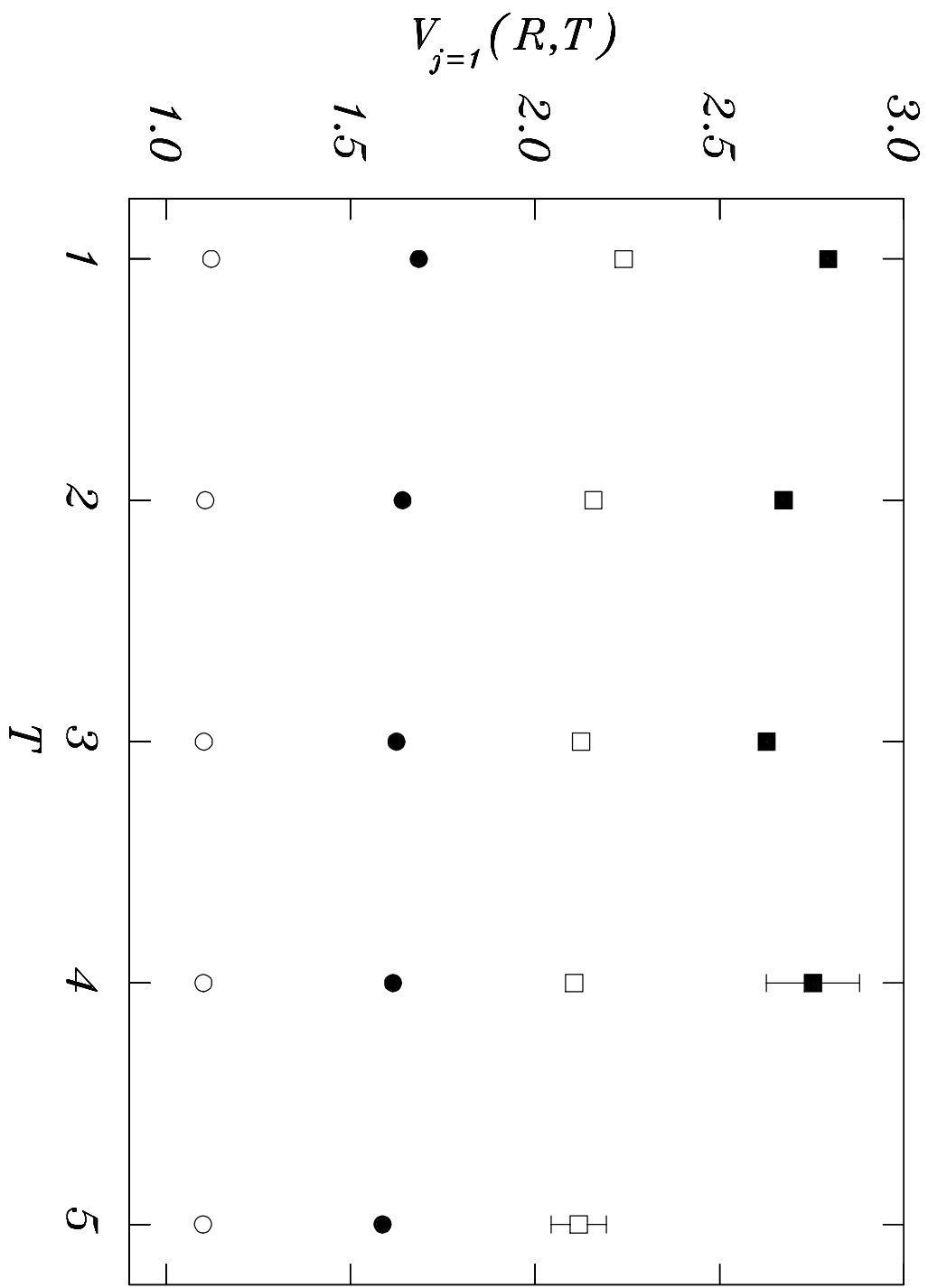

FIGURE 7

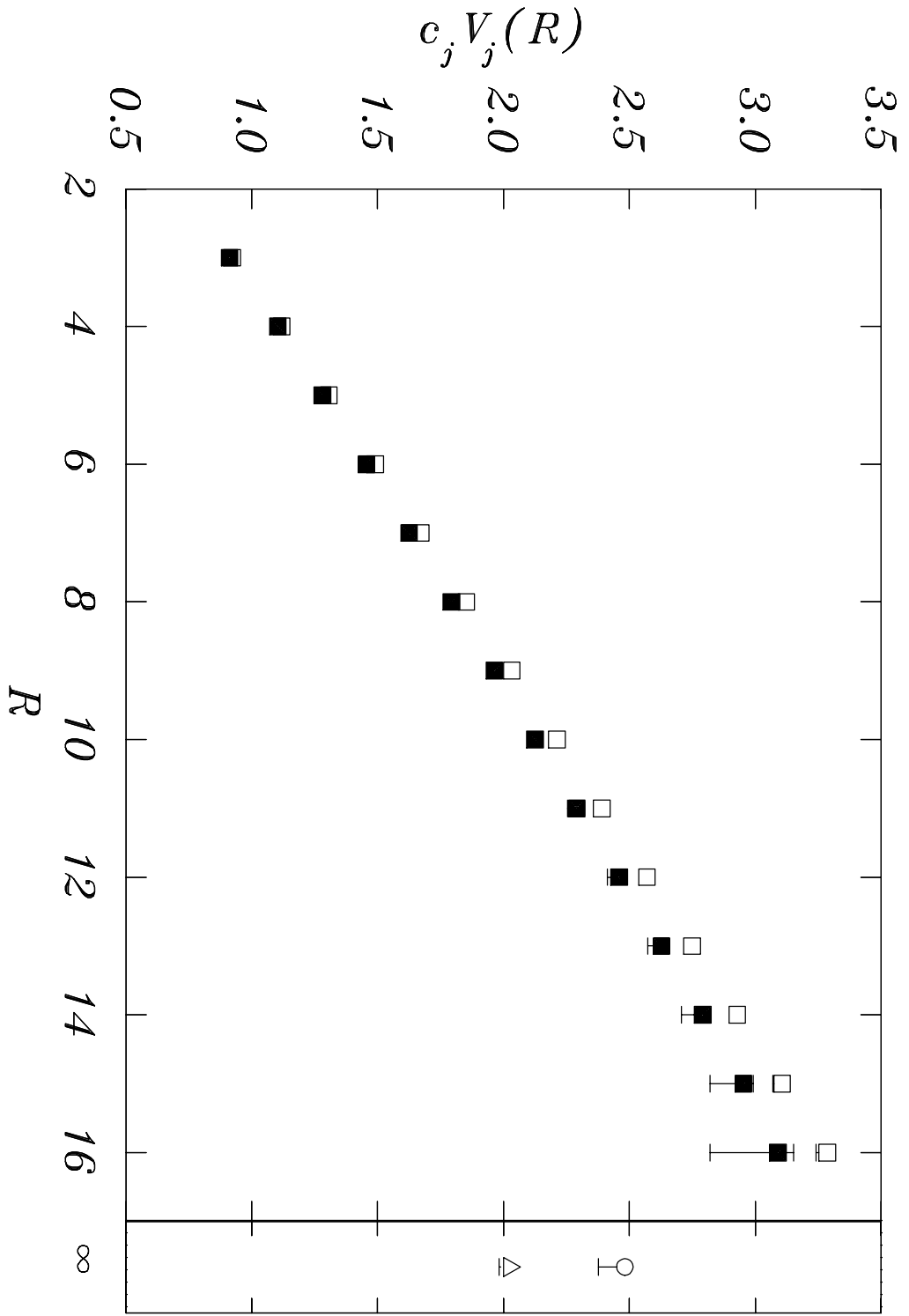
FIGURE 8

# "Gluelump" spectrum and adjoint source potential in lattice QCD$_3$

Grigorios I. Poulis

*NIKHEF-K, Theory Group, Postbus 41882, 1009 DB Amsterdam, The Netherlands*[*]

Howard D. Trottier

*Department of Physics, Simon Fraser University, Burnaby, B.C., Canada V5A 1S6*[†]

(April 1995)

## Abstract

We calculate the potential between "quarks" which are in the adjoint representation of SU(2) color in the three-dimensional lattice theory. We work in the scaling region of the theory and at large quark separations $R$. We also calculate the masses $M_{Qg}$ of color-singlet bound states formed by coupling an adjoint quark to adjoint glue ("gluelumps"). Good scaling behavior is found for the masses of both magnetic (angular momentum $J = 0$) and electric ($J = 1$) gluelumps, and the magnetic gluelump is found to be the lowest-lying state. It is naively expected that the potential for adjoint quarks should saturate above a separation $R_{\rm scr}$ where it becomes energetically favorable to produce a pair of gluelumps. We obtain a good estimate of the naive screening distance $R_{\rm scr}$. However we find little evidence of saturation in the potential out to separations $R$ of about twice $R_{\rm scr}$.

Typeset using REVT$_{\rm E}$X

---

[*]Email address: gregory@nikhefk.nikhef.nl.

[†]Email address: trottier@sfu.ca.

*Introduction* The formation of chromoelectric flux-tubes between static quarks has been well established in lattice QCD simulations. However the physical mechanism underlying confinement remains to be understood, and there continues to be considerable interest in investigations of detailed properties of heavy quark-antiquark ($Q\overline{Q}$) potentials [1] and flux-tubes [2–4]. Of particular relevance both theoretically and phenomenologically is the process of hadronization, or flux-tube breaking. Some evidence for hadronization of a $Q\overline{Q}$ pair has been obtained in nonquenced lattice QCD [4]; the fields in between the quarks were found to be somewhat suppressed in a simulation with dynamical quarks, as compared to the quenched theory. This is consistent with the naive expectation that the heavy quarks should be screened at large $R$, where it becomes energetically favorable for the flux-tube to fission, with the creation of a pair of light valence quarks. However the lattices used in Ref. [4] were too small to demonstrate complete screening of the $Q\overline{Q}$ pair.

Further insight into the physics of confinement and hadronization can be obtained from an analysis of sources which transform in the adjoint representation of the color gauge group [5–9]. The potential between a pair of adjoint "quarks" in SU(2) color has been shown to increase with separation $R$ out to a distance of at least 1.4 fm (where the scale is set by the string tension for fundamental representation quarks) [8]. Flux-tube formation between adjoint ("isospin" $j = 1$) quarks at modest separations has also been demonstrated [9], and the flux-tube was found to have a similar cross-sectional structure as for fundamental representation ($j = 1/2$) quarks.

On the other hand it is expected that a pair of adjoint quarks should not be permanently confined. At large $R$ the quarks should be completely screened, due to the spontaneous creation of valence glue, which binds to each adjoint quark to produce a pair of color-singlet quark-glue bound states ("gluelumps") [6–8]. The energetics of this "hadronization" process should be controlled by the mass $M_{Qg}$ of the lowest-lying gluelump. Naively the potential $V_{j=1}(R)$ between adjoint quarks is expected to saturate at a distance $R_{\text{scr}}$ determined by [6–8]:

$$V_{j=1}(R_{\text{scr}}) \approx 2M_{Qg}. \tag{1}$$

This screening hypothesis is expected to apply to the quenched pure gauge theory, where screening of fundamental representation sources is absent. Results have been obtained for the adjoint quark potential and gluelump masses in SU(2) lattice gauge theory in four dimensions [6–8]. Although the adjoint quark potential is found to deviate appreciably from linearity, results could only be obtained for $R \lesssim R_{\text{scr}}$ [8]; larger separations are required in order to test Eq. (1). Some evidence for adjoint quark screening is also provided by lattice calculations of the chromoelectric field distributions [9].

To shed some light on this question we consider the naive screening hypothesis Eq. (1) in three-dimensional SU(2) lattice gauge theory (QCD$_3$). Lattice QCD$_3$ has proven to be a useful laboratory for the study of the physics of confinement [10–14]. In particular QCD$_3$ is a confining theory exhibiting flux-tube formation and a glueball spectrum. Hence it may be expected that Eq. (1) would apply to QCD in both three and four dimensions.

The first calculation of gluelump masses in QCD$_3$ is presented here [15], and we compute the $Q\overline{Q}$ potential for adjoint quarks to much larger $R$ than has been previously attained [12]. These calculations likewise probe much larger $R$ (compared to $R_{\text{scr}}$) than has been reached in four-dimensional QCD [8]. Good scaling behavior is found for the masses of the magnetic (angular momentum $J = 0$) and electric ($J = 1$) gluelumps, and the magnetic gluelump is found to be the lowest-lying state. Our results provide for a stringent test of Eq. (1). From our calculations we obtain a good estimate of the naive screening distance $R_{\text{scr}}$. However our results for the potential show little evidence of saturation out to separations $R$ of about twice $R_{\text{scr}}$. This represents an interesting challenge to our understanding of the hadronization process, given the basic asumptions that underlie the screening hypothesis of Eq. (1), and merits further investigation.

*Method* Following Refs. [6–8] we extract gluelump masses using an adjoint Polyakov line which is coupled at the two ends to spatial plaquettes. This is accomplished by the following gauge-invariant operator (see Fig. 1):

$$G(T) = \text{Tr}(U_0 \sigma^a) A^{ab}(Q) \text{Tr}(U_T^\dagger \sigma^b), \qquad (2)$$

where $Q$ denotes the Polyakov line of length $T$ in the fundamental representation, which is mapped into the adjoint representation by the operator $A^{ab}$, given by [6]

$$A^{ab}(Q) = \tfrac{1}{2}\text{Tr}\left(\sigma^a Q \sigma^b Q^\dagger\right), \qquad (3)$$

with $\sigma^a$ the Pauli matrices ($a = 1, 2, 3$). $U_0$ and $U_T$ denote linear combinations of fundamental representation spatial plaquettes that share a corner with the lower and upper ends of the Polyakov line, respectively. Using the identity

$$\sigma^a_{ij}\sigma^a_{kl} = 2\left(\delta_{il}\delta_{jk} - \tfrac{1}{2}\delta_{ij}\delta_{kl}\right) \qquad (4)$$

leads to an expression for $G$ that is convenient for computation:

$$G(T) = 2\,\text{Tr}\left(U_0 Q U_T^\dagger Q^\dagger\right) - \text{Tr}(U_0)\text{Tr}(U_T). \qquad (5)$$

For SU(2) we see in Eq. (2) that from the two combinations of plaquettes, e.g. $U_0 \pm U_0^\dagger$, only the "−" contributes [6]; hence only negative charge-parity gluelumps exist.

In order to form an operator $G$ in a definite representation of the cubic group in two spatial dimensions, we use linear combinations of the four oriented spatial plaquettes that share a corner with an end of the Polyakov line. The linear combination in Fig. 2(a) has a symmetry direction perpendicular to the two-dimensional spatial plane, and in the continuum limit should excite a scalar ("magnetic") gluelump. The linear combination in Fig. 2(b) has symmetry direction up and down the page, and is to be idenfitied with a vector-like ("electric") gluelump (another linear combination with symmetry direction across the page would excite the other degenerate vector state). We obtain the gluelump masses $M_{Qg}$ from the $T \to \infty$ extrapolation of the following time-dependent estimates:

$$M_{Qg}(T) = -\ln\left[\frac{G(T)}{G(T-1)}\right]. \qquad (6)$$

A dramatic enhancement in the overlap of $G$ with the lowest-lying state can be achieved by using nonlocal or "fuzzy" plaquettes. Given a set of links $U_\mu(x)$, a corresponding set of "fuzzy" links $\widetilde{U}^1_\mu(x)$ is constructed according to [16]

$$\tilde{U}_{\mu \neq 0}^{1}(x) = \mathcal{N}\left[c\, U_{\mu}(x) + \sum_{\nu \neq \pm\mu, \pm 0} U_{\nu}(x) U_{\mu}(x+\hat{\nu}) U_{\nu}^{\dagger}(x+\hat{\mu})\right], \qquad (7)$$

where $c$ is a positive constant, and $\mathcal{N}$ is an arbitrary normalization (conveniently chosen so that $\det \tilde{U}^1 = 1$). This procedure can be iterated. The number of iterations and the parameter $c$ are chosen "empirically" so as to minimize the $T$ dependence of $M_{Q_g}(T)$. In order to preserve a transfer matrix fuzzing is applied only to spatial links ($\mu \neq 0$) [16]. The fuzzy plaquettes used in Eq. (5) are path ordered products of four fuzzy spatial links.

We also calculate the $Q\overline{Q}$ potential for quarks in both the adjoint ($j=1$) and fundamental ($j=1/2$) representations. The $Q\overline{Q}$ pairs are excited using Wilson loops $W_j$ in the two representations

$$W_j(R,T) \equiv \frac{1}{2j+1} \text{Tr} \left\{ \prod_{l \in L} \mathcal{D}_j[U_l] \right\}, \qquad (8)$$

where $L$ denotes the closed loop, and $\mathcal{D}_j[U_l]$ is an appropriate irreducible representation of the link. Fuzzy spatial links are used in Eq. (8). $W_1$ is computed using relations amongst the group characters, which imply $W_1 = (4W_{1/2}^2 - 1)/3$. The potentials for quarks in the two representations are obtained from an extrapolation of the time-dependent estimates:

$$V_j(R,T) = -\ln\left[\frac{W_j(R,T)}{W_j(R,T-1)}\right]. \qquad (9)$$

A significant enhancement in the signal to noise is obtained by making an analytical integration on the time-like links in the Polyakov line and in the Wilson loop [17]:

$$\int [dU_l] \mathcal{D}_j[U_l] e^{-\beta S} = \frac{I_{2j+1}(\beta k_l)}{I_1(\beta k_l)} \mathcal{D}_j[V_l] \int [dU_l] e^{-\beta S}, \qquad (10)$$

where $S$ is the standard Wilson action, and $k_l V_l$ is equal to the sum of the four "staples" coupling to the link $U_l$ ($\det V_l \equiv 1$ and $k_l \geq 0$).

*Results* Magnetic and electric gluelump operators $G(T)$ were calculated for $T = 1$ to 7 on $40^3$ lattices at $\beta = 6, 8, 10$ and $12$. 800 measurements were made on each lattice. 40 heat bath sweeps were made between measurements, yielding integrated autocorrelations times $\tau_{\text{int}} \lesssim 0.5$. Ten iterations of the fuzzing procedure Eq. (7) were used, with $c = 2.5$. Figures 3 and 4 show results for the time-dependent estimates of the magnetic (angular momentum $J=0$) and electric ($J=1$) gluelump masses, Eq. (6). Statistical errors were estimated using the bootstrap method [18].

In order to estimate the systematic errors in the $T \to \infty$ extrapolation of $M_{Q_g}(T)$ we performed a fit to the transfer matrix, assuming that one excited state dominates the extrapolation [1]:

$$\frac{G(T)}{G(T-1)} \approx \frac{\lambda_0^T + b\lambda_1^T}{\lambda_0^{(T-1)} + b\lambda_1^{(T-1)}}, \qquad (11)$$

where $\lambda_0 = \exp[-M_{Q_g}(T \to \infty)]$ and $\lambda_1/\lambda_0 = \exp(-\Delta M)$ with $\Delta M$ the energy gap to the first excited state. Note that $M_{Q_g}(T)$ at any $T$ provides an upper bound to the true mass.

We fitted Eq. (11) to our data for $G(T)$ with $T$ generally in the range 1–5. Good quality fits were obtained for the data at all four $\beta$ values. To estimate the uncertainty in the fitted values of $\lambda_0$ we performed a bootstrap analysis. We chose configurations at random from the available data (with replacement), and for each such choice we repeated the fit to Eq. (11). The variance over roughly one thousand random selections provided an estimate $\sigma_0$ of the one standard deviation error in $\lambda_0$.

Figure 5 shows our results for the gluelump masses versus $\beta$. In QCD$_3$ (which is super-renormalizable) $\beta = 4/(g^2 a)$, where the coupling constant $g^2$ has dimensions of mass. The gluelump mass in physical units is therefore proportional to $\beta M_J$. The data points in Fig. 5 are upper limits (with statistical errors) given by $M_{Qg}(T)$ at $T = 5$, except at $\beta = 6$ which is taken from $T = 4$. The lower error bars are estimates of the systematic errors in the $T \to \infty$ extrapolation of the masses, determined by $-\ln(\lambda_0 + \sigma_0)$. The estimates of the masses scale within errors, of better than about 10%.

We applied a similar analysis to our data for the Wilson loops. 4,000 measurements were made of Wilson loops of dimension $R \times T = 3 \times 1$ to $9 \times 5$ on a $32^3$ lattice at $\beta = 6$, which is within the scaling region for the potentials [12,14]. 20,000 measurements of Wilson loops with $R = 10$ to $16$ were made on a $40^3$ lattice, also at $\beta = 6$. Between 20 and 40 heat bath updates were made between measurments, again yielding $\tau_{\text{int}} \lesssim 0.5$ (cf. Ref. [12]).

Results for the time-dependent estimates of the potentials, Eq. (9), are shown in Figs. 6 and 7. The fuzzing procedure is evidently very effective at filtering out the $Q\overline{Q}$ ground state. Systematic errors in the $T \to \infty$ extrapolation were estimated by a fit to the transfer matrix, as in Eq. (11), using Wilson loop data with $T$ in the range 1–3. A good check on this procedure is provided by the fact that the fitted functions agree with the Wilson loop data for larger $T$ within one or two standard deviations (however useful results for $R \geq 14$ could only be obtained for $T \leq 3$).

Figure 8 shows our results for the potentials as functions of $R$. The potentials are also compared with twice the gluelump masses, shown in the right-hand box in Fig. 8. The plotted points for the potential are upper limits (with statistical errors) given by the data at $T = 3$. The lower error bars are the estimates of the systematic errors. The data for the $j = 1/2$ quarks have been rescaled by the ratio of the Casimirs $j(j + 1)$ of the two representations, equal to $8/3$.

The fact that $V_{j=1}(R)$ falls below $8/3 \times V_{j=1/2}(R)$ at large $R$ can be interpreted as evidence for partial screening of the adjoint quarks (the Casimir can be interpreted as the squared "charge" of the quarks) [6–9]. However the screening is much less than is suggested by Eq. (1), which implies a screening distance $R_{\text{scr}}$ of about 9 lattice units, based on the lightest (magnetic) gluelump mass. This is to be compared with the fact that at $R = 16$ the adjoint potential potential $V_{j=1}$ is still within about 10% of $8/3 \times V_{j=1/2}$.

This represents an interesting challenge to our understanding of the hadronization process, given the basic assumptions that underlie Eq. (1), and merits further investigation. We note that the calculations done here in QCD$_3$ probed much larger $R$ (compared to $R_{\text{scr}}$) than has been reached in the four-dimensional theory [8]; hence there is as yet no evidence for saturation of the adjoint quark potential in either theory. Some clarification of this problem might be obtained in QCD$_3$ from an analysis of the gluelump spatial structure, the interaction between a pair of gluelumps, and their mixing with a $Q\overline{Q}$ pair (cf. Ref. [6–8]). We are currently investigating these possibilities.

We thank Richard Woloshyn, Jan Smit and Piet Mulders for fruitful discussions. The work of H.D.T. was supported in part by the Natural Sciences and Engineering Research Council of Canada. The work of G.I.P. was supported in part by Human Capital & Mobility EC Fellowship ERBCHBICT941430.

# REFERENCES

[1] See, e.g., S. P. Booth et al., Nucl. Phys. B394 (1993) 509.
[2] R. W. Haymaker and J. Wosiek, Phys. Rev. D 36 (1987) 3297; Acta Phys. Pol. B21 (1990) 403; Phys. Rev. D 43 (1991) 2676.
[3] M. Fukugita and T. Niuya, Phys. Lett. B132 (1983) 374; J. W. Flower and S. W. Otto, ibid., B160 (1985) 128; R. Sommer, Nucl. Phys. B291 (1986) 673; A. Di Giacomo, M. Maggiore, and S. Olejnik, Phys. Lett. B236 (1990) 199; G.S. Bali, K. Schilling, and Ch. Schlichter CERN preprint CERN-TH 7413/94 (1994).
[4] W. Feilmair and H. Markum, Nucl. Phys. B 370 (1992) 299; W. Bürger et al., Nucl. Phys. B (Proc. Suppl.) 20 (1991) 203.
[5] C. Bernard, Phys. Lett. B108 (1982) 431; J. Ambjørn, P. Olesen, and C. Peterson, Nucl. Phys. B240 (1984) 189.
[6] C. Michael, Nucl. Phys. B 259 (1985) 58.
[7] L. A. Griffiths, C. Michael, and P. Rakow, Phys. Lett. B150 (1985) 196; N. A. Campbell, I. H. Jorysz, and C. Michael, ibid., B167 (1986) 91; I. H. Jorysz and C. Michael, Nucl. Phys. B302 (1987) 448.
[8] C. Michael, Nucl. Phys. B (Proc. Suppl.) 26 (1992) 417.
[9] H. D. Trottier, SFU preprint HEP-115-95, 1995.
[10] J. Ambjørn, P. Olesen and C. Peterson, Nucl. Phys. B240 (1984) 533.
[11] D. G. Caldi and T. Sterling, Phys. Rev. Lett. 60 (1988) 2454;
[12] R.D. Mawhinney, Phys. Rev. D 41 (1990) 3209.
[13] M. Teper, Phys. Lett. B 289 (1992) 115.
[14] H. D. Trottier and R. M. Woloshyn, Phys. Rev. D 48 (1993) 2290.
[15] Preliminary results were reported in G. I. Poulis and H. D. Trottier, Proceedings of the 12th International Symposium on Lattice Field Theory, Bielefeld, Germany, 27 Sep - 1 Oct 1994.
[16] M. Teper, Phys. Lett. B183 (1987) 345; M. Albanese et al., Phys. Lett. B192 (1987) 163.
[17] G. Parisi, R. Petronzio and F. Rapuano, Phys. Lett. B 126(1983) 250. See also Ref. [12].
[18] See e.g. B. Efrom, SIAM Review 21 (1979) 460.

FIGURES

FIG. 1. The gluelump operator $G(T)$. The double line represents the adjoint Polyakov line.

FIG. 2. The combination of spatial plaquettes for (a) the magnetic gluelump and (b) the electric gluelump. The filled circle represents the position of the Polyakov line.

FIG. 3. Time-dependent magnetic gluelump mass $M_{J=0}(T,\beta)$ for $\beta = 6$ ($\circ$), 8 ($\bullet$), 10 ($\square$), and 12 ($\blacksquare$).

FIG. 4. Time-dependent electric gluelump mass $M_{J=1}(T,\beta)$. The plotted symbols are the same as in Fig. 3.

FIG. 5. Magnetic ($\square$) and electric ($\blacksquare$) gluelump masses versus $\beta$.

FIG. 6. Time-dependent potential $V_{j=1/2}(R,T)$ for fundamental representation quarks for $R = 4$ ($\circ$), 8 ($\bullet$), 12 ($\square$) and 16 ($\blacksquare$).

FIG. 7. Time-dependent potential $V_{j=1}(R,T)$ for adjoint representation quarks for $R = 4$ ($\circ$), 7 ($\bullet$), 10 ($\square$) and 13 ($\blacksquare$).

FIG. 8. $Q\overline{Q}$ potentials versus separation $R$ for $j = 1/2$ ($\square$) and $j = 1$ ($\blacksquare$). The $j = 1/2$ potential has been rescaled by $c_{1/2} \equiv 8/3$ ($c_1 \equiv 1$). The box on the right shows twice the magnetic ($\triangle$) and electric ($\circ$) gluelump masses.